\documentclass[12pt]{article}

\begin{document}
\begin{center}
 {\bf{\Large B\"{a}cklund Transformations of Einstein's Field Equations for
  the Interior of a Uniformly Rotating Stationary Axisymmetric Perfect
  Fluid}}
\vspace{1cm}

E. Kyriakopoulos\\ Department of Physics\\ National Technical
University\\ 157 80 Zografou, Athens, GREECE
\end{center}

\begin{abstract}
Clairin's method of obtaining B\"{a}cklund transformations is
applied to Einstein's field equations for the interior of a
uniformly rotating stationary axisymmetric perfect fluid. It is
shown that for arbitrary pressure $ p $ and mass density $ \mu $
the method does not give non-trivial B\"{a}cklund transformations,
while if $\mu + 3p =0 $ it gives the transformation of Ehlers.

PACS number(s): 04.20.Jb, 04.20.Ex
\end{abstract}

\section{Introduction}
   To get solutions of a non-linear second order partial differential
equation or a system of such equations one may use B\"{a}cklund
Transformations (BTs), provided that these equations do have such
transformations \cite{M1}.A BT consists of a set of first order
partial differential equations relating a solution of the given
equations to another solution of the given equations ( or to a
solution of another set of equations.) A method of getting BTs has
been introduced by Clairin \cite{C2}, \cite{C3}  and it has been
shown that by this method we get all known BTs of the Ernst's
equation \cite{O4} i.e. the transformations of Ehlers \cite{E5} of
Neugenbauer \cite{N6} and of Harrison \cite{H7}. Also by this
method a BT of the Einstein-Maxwell equations has been found
\cite{O8}.

BTs of Einstein's equations for the interior of a uniformly
rotating, stationary, axisymmetric perfect fluid are not known and
this is one of the reasons the number of solutions of these
equations is limited. In this work we shall try to find BTs of
these equations by Clairin's method. We use a form of Einstein's
equations which is very convenient for the application of the
method \cite{K9}. In Sect 2 we apply the method in the general
case i.e. in the case in which we don't impose any specific
relation between the pressure $ p $ and the energy density $ \mu $
of the fluid i.e. for arbitrary equation of state. We show that in
this case the method does not give no-trivial BTs. In Sect. 3 we
assume for the fluid an equation of state of the form $ p+3 \mu =
0 $ and we get non-trivial BTs. The general solution of the system
of first order partial differential equations of the BTs is found,
and it is shown that this leads to the transformation of Ehlers
\cite{E5}.
\section{B\"{a}cklund Transformations in the General Case}
We shall give some definitions and briefly state some results of
Ref \cite{K9}. The metric for stationary axially symmetric space
times admitting 2-spaces orthogonal to the Killing vectors
$\vec{\xi} =\partial_t$ and  $\vec{\eta} =\partial_\phi$ can be
 written in the form
\begin{equation}
ds^{2}=e^{-2U}\{e^{2K}(d\rho^{2}+dz^{2})+
F^{2}d\phi^{2}\}-e^{2U}(dt+Ad\phi)^{2}   \label{21}
\end{equation}
where $U=U(\rho,z)$, $K=K(\rho,z)$, $F=F(\rho,z)$ and
$A=A(\rho,z)$. For a perfect fluid source we have the
energy-momentum tensor
\begin{equation}
 T_{ab}=(\mu+p)u_{a}u_{b}+pg_{ab}  \label{22}
\end{equation}
where $u_{a}$ are the components of the 4-velocity and $\mu$ and $
p $ are the  mass density and the pressure of the fluid
respectively.

We shall introduce the notation
\begin{eqnarray}
A=-\omega \mbox{, } e^{2U}=T  \mbox{, } \xi=\rho+iz \mbox{, }
\eta=\rho-iz  \label{23}
\end{eqnarray}
and the potentials $ \phi $ and $ E $ by
\begin{eqnarray}
\omega_{\rho}=\frac{F}{T^{2}}\phi_z \mbox{,  }
\omega_z=-\frac{F}{T^{2}}\phi_\rho  \mbox{, } E=T+i\phi \label{24}
\end{eqnarray}
where a letter as an index in a function means differentiation
with respect to the corresponding variable e.g.
$\omega_{\rho}=\frac{\partial \omega}{\partial \rho}$. Also we put
\begin{eqnarray}
&&lnF=G \mbox{, } 2K-G=M  \label{25}\\
&&\Lambda=\frac{1}{2}(E+\overline{E})(\nabla^{2}E
+\frac{1}{F}\vec{\nabla}F\cdot\vec{\nabla}E)
-\vec{\nabla}E\cdot\vec{\nabla}E=0  \label{26}
\end{eqnarray}
where
\begin{eqnarray}
\nabla^{2}=\partial^{2}_{\rho\rho}+\partial^{2}_{zz} \mbox{, }
\vec{\nabla}=\hat{\rho}\partial_{\rho}+\hat{z}\partial_{z}
\label{27}
\end{eqnarray}
with $\hat{\rho} $ and $\hat{z}$ unit vectors in the direction of the
$\rho$ and $ z $ axis respectively. Then Einstein's field equations take
the form \cite{K9}
\begin{eqnarray}
&&G_{\eta\eta}-M_{\eta}G_{\eta}+\frac{1}{2 T^{2}}(T^{2}_{\eta}+
\phi^{2}_{\eta})=0 \label{28}\\
&&M_{\eta\xi}-G_{\eta}G_{\xi}+\frac{1}{2T^{2}}(T_{\eta}T_{\xi}
+\phi_{\eta}\phi_{\xi})=0\label{29}\\
&&\frac{1}{T}Im\Lambda=\phi_{\eta\xi}+\frac{1}{2}(G_{\eta}\phi_{\xi}
+G_{\xi}\phi_{\eta})-\frac{1}{T}(T_{\eta}\phi_{\xi}+T_{\xi}\phi_{\eta})=0
\label{210}\\
&&p=2Te^{-G-M}(G_{\eta\xi}+G_{\eta}G_{\xi}) \label{211}\\
&&\mu=-3p+\frac{1}{T}Re\Lambda \label{212}
\end{eqnarray}
where the expressions $ Im\Lambda $ and $ Re\Lambda $ are the
imaginary and the real parts of  $ \Lambda $ respectively.

We shall try to find BTs in the general case i.e. if we don't
impose any relation between $ p $ and $ \mu $. We shall apply the
method of Clairin \cite{C2}, \cite{C3}  to Eqs (\ref{28}) -
(\ref{210}). Having found such a transformation we shall use Eqs
(\ref{211}) and (\ref{212}) to determine $ p' $ and $ \mu' $ of
the new solution.The most general forms of BTs are :
\begin{eqnarray}
&&G'_{\eta} = \alpha_{0}+\alpha_{1}G_{\eta}+\alpha_{2}M_{\eta}
+\alpha_{3}T_{\eta}+\alpha_{4}\phi_{\eta}+\alpha_{5}G_{\xi}\nonumber\\
&&+\alpha_{6}M_{\xi}+\alpha_{7}T_{\xi}+\alpha_{8}\phi_{\xi}
\label{213}\\
&&M'_{\eta} = \beta_{0}+\beta_{1}G_{\eta}+\beta_{2}M_{\eta}+
\beta_{3}T_{\eta}+
\beta_{4}\phi_{\eta}+\beta_{5}G_{\xi}\nonumber\\
&&+\beta_{6}M_{\xi}+\beta_{7}T_{\xi} + \beta_{8}\phi_{\xi}
\label{214}\\
&&T'_{\eta}=\gamma_{0}+\gamma_{1}G_{\eta}+\gamma_{2}M_{\eta}
+\gamma_{3}T_{\eta}+\gamma_{4}\phi_{\eta}+\gamma_{5}G_{\xi}\nonumber\\
&&+\gamma_{6}M_{\xi}+\gamma_{7}T_{\xi}+\gamma_{8}\phi_{\xi}
\label{215}\\
&&\phi'_{\eta}=\delta_{0}+\delta_{1}G_{\eta}+ \delta_{2}M_{\eta}
+\delta_{3}T_{\eta}+\delta_{4}\phi_{\eta}+\delta_{5}G_{\xi}\nonumber\\
&&+\delta_{6}M_{\xi}+\delta_{7}T_{\xi}+\delta_{8}\phi_{\xi}
\label{216}\\
&&G'_{\xi}=\overline{G'_{\eta}} \mbox{, }
M'_{\xi}=\overline{M'_{\eta}} \mbox{, }
T'_{\xi}=\overline{T'_{\eta}} \mbox{, }
\phi_{\xi}=\overline{\phi'_{\eta}} \label{217}
\end{eqnarray}
where $ R=R(G,M,T,\phi,G',M',T',\phi',\xi,\eta) $, $ R = \alpha_{0},
\alpha_{1}, ...,\alpha_{8}, \beta_{0},...,\delta_{8} $. If we
substitute the above expressions into Eqs (\ref{28})-(\ref{210})
for the quantities with prime and into the integrability
conditions
\begin{equation}
G'_{\eta\xi}-G'_{\xi\eta}=M'_{\eta\xi}-M'_{\xi\eta}=T'_{\eta\xi}-T'_{\xi\eta}
=\phi'_{\eta\xi}-\phi'_{\xi\eta}=0 \label{218}
\end{equation}
the relations we shall get should hold for any solution. Therefore
the coefficients of all linearly independent quantities should
vanish. Such quantities are : $ G_{\eta\xi}$, $ M_{\eta\eta}$, $
M_{\xi\xi}$, $ T_{\eta\xi}$, $ T_{\eta\eta}$, $ T_{\xi\xi}$, $
\phi_{\eta\eta}$, $ \phi_{\xi\xi}$, $ S_{i}S_{j}$, $ S_{i}$, where
$ S $ = $ G $, $ M $, $ T $, $ \phi $, and i, j = $ \xi $, $ \eta
$.

We get
\begin{eqnarray}
&&G'_{\eta\eta}-M'_{\eta}G'_{\eta}+\frac{1}{2T'^{2}}(T'^{2}_{\eta}
+\phi'^{2}_{\eta})=\alpha_{2}M_{\eta\eta}+\alpha_{3}T_{\eta\eta}\nonumber\\
&&+\alpha_{4}\phi_{\eta\eta}+\alpha_{5}G_{\xi\eta}+\alpha_{7}T_{\xi\eta}+...=0
\label{219}\\
&&G'_{\eta\xi}-G'_{\xi\eta}=(\alpha_{1}-\overline{\alpha_{1}})G_{\eta\xi}
+\alpha_{6}M_{\xi\xi}+\alpha_{8}\phi_{\xi\xi}+...=0
\label{220}\\
&&M'_{\eta\xi}-G'_{\eta}G'_{\xi}+\frac{1}{2T'^{2}}(T'_{\xi}T'_{\eta}
+\phi'_{\xi}\phi'_{\eta})=\beta_{1}G_{\eta\xi}+\beta_{3}T_{\eta\xi}\nonumber\\
&&+\beta_{6}M_{\xi\xi}+\beta_{7}T_{\xi\xi}+\beta_{8}\phi_{\xi\xi}+...=0
\label{221}\\
&&T'_{\eta\xi}-T'_{\xi\eta}=(\gamma_{1}-\overline{\gamma_{1}})G_{\eta\xi}+
(\gamma_{3}-\overline{\gamma_{3}})T_{\eta\xi}+\gamma_{6}M_{\xi\xi}\nonumber\\
&&+\gamma_{7}T_{\xi\xi}+\gamma_{8}\phi_{\xi\xi}+...=0
\label{222}\\
&&\phi'_{\eta\xi}+\frac{1}{2}(G'_{\eta}\phi'_{\xi}+G'_{\xi}\phi'_{\eta})
-\frac{1}{T'}(T'_{\xi}\phi'_{\eta}+T'_{\eta}\phi'_{\xi})\nonumber\\
&&=\delta_{1}G_{\eta\xi}+\delta_{3}T_{\eta\xi}+\delta_{6}M_{\xi\xi}
+\delta_{7}T_{\xi\xi}+\delta_{8}\phi_{\xi\xi}+...=0
\label{223}
\end{eqnarray}
where only some of the linearly independent terms with their
coefficients are shown.Therefore we must have
\begin{eqnarray}
&&\alpha_{2}=\alpha_{3}=\alpha_{4}=\alpha_{5}=\alpha_{6}=\alpha_{7}
=\alpha_{8}=0 \label{224}\\
&&\overline{\alpha_{1}}=\alpha_{1} \label{225}\\
&&\beta_{1}=\beta_{3}=\beta_{6}=\beta_{7}=\beta_{8}=0 \label{226}\\
&&\gamma_{6}=\gamma_{7}=\gamma_{8}=0  \label{227}\\
&&\overline{\gamma_{1}}=\gamma_{1} \mbox{, } \overline{\gamma_{3}}
=\gamma_{3} \label{228}\\
&&\delta_{1}=\delta_{3}=\delta_{6}=\delta_{7}=\delta_{8}=0
\label{229}
\end{eqnarray}
and the BTs become
\begin{eqnarray}
&&G'_{\eta}=\alpha_{1}G_{\eta}+\alpha_{0} \label{230} \\
&&M'_{\eta}=\beta_{2}M_{\eta}+\beta_{4}\phi_{\eta}+\beta_{5}G_{\xi}
+\beta_{0} \label{231}\\
&&T'_{\eta}=\gamma_{1}G_{\eta}+\gamma_{2}M_{\eta}+\gamma_{3}T_{\eta}
+\gamma_{4}\phi_{\eta}+\gamma_{5}G_{\xi}+\gamma_{0} \label{232}\\
&&\phi'_{\eta}=\delta_{2}M_{\eta}+\delta_{4}\phi_{\eta}
+\delta_{5}G_{\xi}+\delta_{0} \label{233}
\end{eqnarray}
Also using expressions (\ref{230})-(\ref{233}) and the equations
which the functions G, M, and $\phi $ satisfy we get :
\begin{eqnarray}
&&G'_{\eta\eta}-M'_{\eta}G'_{\eta}
+\frac{1}{2T'^{2}}(T'^{2}_{\eta}+\phi'^{2}_{\eta})
=\frac{1}{2T'^{2}}(\gamma^{2}_{2}+\delta^{2}_{2})M^{2}_{\eta}+...=0
\label{234}\\
&&M'_{\eta\xi}-M'_{\xi\eta}=(\beta_{2}-\overline{\beta_{2}})G_{\eta}G_{\xi}
-\frac{\beta_{5}}{2T^{2}}(T^{2}_{\xi}+\phi^{2}_{\xi})+...=0
\label{235}\\
&&T'_{\eta\xi}-T'_{\xi\eta}=-\frac{\gamma_{5}}{2T^{2}}(T^{2}_{\xi}
+\phi^{2}_{\xi})+...=0 \label{236}\\
&&\phi'_{\eta\xi}+\frac{1}{2}(G'_{\eta}\phi'_{\xi}+G'_{\xi}\phi'_{\eta})
-\frac{1}{T'}(T'_{\xi}\phi'_{\eta}+T'_{\eta}\phi'_{\xi})\nonumber\\
&&=\{\delta_{2}-\frac{1}{T'}(\gamma_{5}\overline{\delta_{5}}
+\overline{\gamma_{5}}\delta_{5})\}G_{\xi}G_{\eta}
-\frac{\delta_{5}}{2T^{2}}(T^{2}_{\xi}+\phi^{2}_{\xi})+...=0
\label{237}
\end{eqnarray}
Therefore we must have
\begin{eqnarray}
&&\beta_{5}=\gamma_{2}=\gamma_{5}=\delta_{2}=\delta_{5}=0 \label{238}\\
&&\overline{\beta_{2}}=\beta_{2}  \label{239}
\end{eqnarray}
Thus the BTs become
\begin{eqnarray}
&&G'_{\eta}=\alpha_{1}G_{\eta}+\alpha_{0} \label{240}\\
&&M'_{\eta}=\beta_{2}M_{\eta}+\beta_{4}\phi_{\eta}+\beta_{0}\label{241}\\
&&T'_{\eta}=\gamma_{1}G_{\eta}+\gamma_{3}T_{\eta}
+\gamma_{4}\phi_{\eta}+\gamma_{0} \label{242}\\
&&\phi'_{\eta}=\delta_{4}\phi_{\eta}+\delta_{0} \label{243}
\end{eqnarray}
where $ \alpha_{1} $, $\beta_{2}$, $\gamma_{1}$ and $\gamma_{3}$ are real
functions.

To proceed further we introduce the notation
\begin{eqnarray}
&&\Delta_{,\:G}\equiv\Delta_{G}+\alpha_{1}\Delta_{G'}
+\gamma_{1}\Delta_{T'} \nonumber\\
&&\Delta_{,\:M}\equiv\Delta_{M}+\beta_{2}\Delta_{M'}\nonumber\\
&&\Delta_{,\:T}\equiv\Delta_{T}+\gamma_{3}\Delta_{T'}\label{244}\\
&&\Delta_{,\:\phi}\equiv\Delta_{\phi}+\beta_{4}\Delta_{M'}
+\gamma_{4}\Delta_{T'}+\delta_{4}\Delta_{\phi'}\nonumber\\
&&\Delta_{,\:\overline{\phi}}\equiv\Delta_{\phi}
+\overline{\beta_{4}}\Delta_{M'}+\overline{\gamma_{4}}\Delta_{T'}
+\overline{\delta_{4}}\Delta_{\phi'}\nonumber\\
&&\Delta_{,\:\eta}\equiv\Delta_{\eta}+\alpha_{0}\Delta_{G'}
+\beta_{0}\Delta_{M'}+\gamma_{0}\Delta_{T'}+\delta_{0}\Delta_{\phi'}
\nonumber
\end{eqnarray}
where $ \Delta=\alpha_{1} \mbox{, } \alpha_{0} \mbox{,
}\beta_{2}$, ... $ \delta_{0}$.Then using Eqs (\ref{240}) -
(\ref{243}) we get from the equations the functions $ G' \mbox{, } M'$
and $ \phi'$ must satisfy and from the integrability conditions a big number
of equations. We give some of them. From (\ref{28}) for the new
solution we get :
\begin{eqnarray}
&&-\frac{\alpha_{1}}{T^{2}}+\frac{\gamma^{2}_{3}}{T'^{2}}=0 \label{245}\\
&&\alpha_{1,\; \eta}+\alpha_{0 ,\; G}-\alpha_{1}\beta_{0}
+\frac{\gamma_{1}\gamma_{0}}{T'^{2}}=0 \label{246}\\
&&\alpha_{1,\; G}+\frac{\gamma^{2}_{1}}{2T'^{2}}=0 \label{247}\\
&&\gamma_{3}\gamma_{4}=0 \label{248}\\
&&-\frac{\alpha_{1}}{T^{2}}+\frac{\gamma^{2}_{4}
+\delta^{2}_{4}}{T'^{2}}=0 \label{249}
\end{eqnarray}
Also from Eq (\ref{29}) we get :
\begin{eqnarray}
&&\beta_{2}-\alpha^{2}_{1}+\frac{\gamma^{2}_{1}}{2T'^{2}}=0 \label{250}\\
&&\gamma_{1}\gamma_{3}=0 \label{251} \\
&&-\frac{\beta_{4}}{2}+\frac{\gamma_{1}\overline{\gamma_{4}}}{2T'^{2}}=0
\label{252}\\
&&-\alpha_{1}\overline{\alpha_{0}}
+\frac{\gamma_{1}\overline{\gamma_{0}}}{T'^{2}}=0 \label{253}\\
&&\beta_{2,\; \xi}=0 \label{254} \\
&&-\frac{\beta_{2}}{T^{2}}+\frac{\gamma^{2}_{3}}{T'^{2}}=0 \label{255}\\
&&\gamma_{3}\overline{\gamma_{0}}=0 \label{256}
\end{eqnarray}
and from Eq (\ref{210}) we get :
\begin{eqnarray}
&&\frac{\delta_{4}}{T}-\frac{\gamma_{3}\overline{\delta_{4}}}{T'}=0
\label{257}\\
&&\gamma_{3}\overline{\delta_{0}}=0 \label{258}
\end{eqnarray}

Eqs (\ref{245})-(\ref{258}) imply that most of the coefficients in
the BTs (\ref{240})-(\ref{243}) vanish. To show that we shall
consider two cases.

Case 1
\begin{equation}
\gamma_{3}\neq0 \label{259}
\end{equation}
Then Eqs (\ref{248}), (\ref{251}), (\ref{252}), (\ref{256}) and
(\ref{258}) imply
\begin{equation}
\beta_{4}=\gamma_{1}=\gamma_{4}=\gamma_{0}=\delta_{0}=0 \label{260}
\end{equation}
But if $ \gamma_{0}=0 $ Eq (\ref{253}) implies the relation
$ \alpha_{1}\alpha_{0}=0.$ However if $ \alpha_{1}=0 $ (\ref{250})
gives $ \beta_{2}=0 $ in which case Eq (\ref{255}) implies the relation
$ \gamma_{3}=0 $, contrary to Eq (\ref{259}). Therefore we must take
\begin{equation}
\alpha_{0}=0 \label{261}
\end{equation}
in which case Eq (\ref{246}) gives $ \alpha_{1,\; \eta}
-\alpha_{1}\beta_{0}=0 $. From Eqs (\ref{225}) and (\ref{250}) we
get $ \beta_{2}=\alpha^{2}_{1} $ : real. Then Eq (\ref{254}) gives
$ \beta_{2,\; \eta}=0 $ and since $ \beta_{2,\; \eta}
=2\alpha_{1}\alpha_{1,\; \eta} $  we find that
$ \alpha_{1,\; \eta}=0 $ since $ \alpha_{1}\neq0 $. Thus we have
\begin{equation}
\beta_{0}=0 \label{262}
\end{equation}
Therefore we get the BT
\begin{equation}
G'_{\eta}=\alpha_{1}G_{\eta},\; M'_{\eta}=\beta_{2}M_{\eta},\;
T'_{\eta}=\gamma_{3}T_{\eta},\; \phi'_{\eta}=\delta_{4}\phi_{\eta}
\label{263}
\end{equation}
where $ \alpha_{1} $, $ \beta_{2} $ and $ \gamma_{3} $ are real
functions. Substituting expressions (\ref{263}) to equations which
$ G' $, $ M' $ and $ \phi' $ satisfy and to the integrability
conditions and solving the resulting equations we get the
trivial relations
\begin{equation}
G'=G+c_{1} \mbox{, } M'=M+c_{2} \mbox{, } T'=c_{3}T \mbox{, }
\phi'=\pm c_{3}\phi+c_{4} \label{264}
\end{equation}
where $ c_{1} $, $ c_{2} $, $ c_{3} $ and $ c_{4} $ are arbitrary
real constants.

Case 2
\begin{equation}
\gamma_{3}=0
\end{equation}
Then from  Eqs (\ref{245}), (\ref{247}), (\ref{249}), (\ref{250}),
(\ref{252}) and (\ref{257}) we get
\begin{equation}
\alpha_{1}=\beta_{2}=\beta_{4}=\gamma_{1}=\gamma_{4}=\delta_{4}=0 \label{266}
\end{equation}
Therefore we obtain the non-interesting relations
\begin{equation}
G'_{\eta}=\alpha_{0},\; M'_{\eta}=\beta_{0},\; T'_{\eta}=\gamma_{0},\;
\phi'_{\eta}=\delta_{0} \label{267}
\end{equation}
Our conclusion is that Clairin's method, which in the vacuum case
gives all known BTs \cite{O4}, does not give interesting BTs in the
general case.
\section{B\"{a}cklund Transformations if $ \mu+3p=0 $ }
If $ \mu+3p=0 $ Eq (\ref{212}) gives $ Re\Lambda=0 $, which
combined with (\ref{210}) gives $ \Lambda=0 $. Therefore the
system of Eqs (\ref{28}) - (\ref{211}) in the variables $ G,\;
M,\;E \mbox{,  and  }  \overline{E} $ (~ for simplicity we write also $ E
+\overline{E}=2T $ ) takes the form
\begin{eqnarray}
&&G_{\eta\eta}-M_{\eta}G_{\eta}+\frac{1}{2 T^{2}}E_{\eta}\overline{E}_{\eta}=0
 \label{31}\\
&&M_{\eta\xi}-G_{\eta}G_{\xi}+\frac{1}{4T^{2}}(E_{\xi}\overline{E}_{\eta}
+\overline{E}_{\xi}E_{\eta})=0 \label{32}\\
&&\frac{1}{T}\Lambda=E_{\eta\xi}+\frac{1}{2}(G_{\eta}E_{\xi}
+G_{\xi}E_{\eta})-\frac{1}{T}E_{\xi}E_{\eta}=0 \label{33}\\
&&p=2Te^{-G-M}(G_{\eta\xi}+G_{\eta}G_{\xi}) \label{34}
\end{eqnarray}

We want to find BTs for the system of Eqs (\ref{31}) - (\ref{33})
following Clairin's method. The most general BTs are given by the
relations
\begin{eqnarray}
&&G'_{\eta}=\zeta_{0}+\zeta_{1}G_{\eta}+\zeta_{2}M_{\eta}+\zeta_{3}E_{\eta}
+\zeta_{4}\overline{E}_{\eta}+\zeta_{5}G_{\xi}\nonumber\\&&+\zeta_{6}M_{\xi}
+\zeta_{7}E_{\xi}+\zeta_{8}\overline{E}_{\xi} \label{35}\\
&&M'_{\eta}=\lambda_{0}+\lambda_{1}G_{\eta}+\lambda_{2}M_{\eta}
+\lambda_{3}E_{\eta}+\lambda_{4}\overline{E}_{\eta}+\lambda_{5}G_{\xi}\nonumber\\
&&+\lambda_{6}M_{\xi}+\lambda_{7}E_{\xi}+\lambda_{8}\overline{E}_{\xi}
\label{36}\\
&&E'_{\eta}=\mu_{0}+\mu_{1}G_{\eta}+\mu_{2}M_{\eta}+\mu_{3}E_{\eta}
+\mu_{4}\overline{E}_{\eta}+\mu_{5}G_{\xi}\nonumber\\&&+\mu_{6}M_{\xi}
+\mu_{7}E_{\xi}+\mu_{8}\overline{E}_{\xi} \label{37}\\
&&\overline{E'}_{\eta}=\nu_{0}+\nu_{1}G_{\eta}+\nu_{2}M_{\eta}+\nu_{3}E_{\eta}
+\nu_{4}\overline{E}_{\eta}+\nu_{5}G_{\xi}\nonumber\\&&+\nu_{6}M_{\xi}+\nu_{7}E_{\xi}
+\nu_{8}\overline{E}_{\xi} \label{38}\\
&&G'_{\xi}=\overline{G'_{\eta}},\; M'_{\xi}=\overline{M'_{\eta}},\;
E'_{\xi}=\overline{\overline{E'}_{\eta}},\;
\overline{E'}_{\xi}=\overline{E'_{\eta}} \label{39}
\end{eqnarray}
while $ p' $ is given by Eq (\ref{34}) in terms of $ T' $, $ G' $, and $ M' $.
Proceeding as in the previous case we find that
\begin{eqnarray}
\zeta_{2}=\zeta_{3}=\zeta_{4}=\zeta_{5}=\zeta_{6}=\zeta_{7}=\zeta_{8}=0,\;
\overline{\zeta_{1}}=\zeta_{1} \label{310}\\
\kappa_{1}=\kappa_{5}=\kappa_{6}=\kappa_{7}=\kappa_{8}=0,\;
\kappa=\lambda,\; \mu,\; \nu,\; \label{311}
\end{eqnarray}Therefore our BTs are reduced to
\begin{eqnarray}
&&G'_{\eta}=\zeta_{0}+\zeta_{1}G_{\eta},\; \overline{\zeta_{1}}=\zeta_{1}
\label{312}\\
&&M'_{\eta}=\lambda_{0}+\lambda_{2}M_{\eta}+\lambda_{3}E_{\eta}
+\lambda_{4}\overline{E}_{\eta} \label{313}\\
&&E'_{\eta}=\mu_{0}+\mu_{2}M_{\eta}+\mu_{3}E_{\eta}+\mu_{4}\overline{E}_{\eta}
\label{314}\\
&&\overline{E'}_{\eta}=\nu_{0}+\nu_{2}M_{\eta}+\nu_{3}E_{\eta}
+\nu_{4}\overline{E}_{\eta} \label{315}
\end{eqnarray}
Then we get
\begin{eqnarray}
&&G'_{\eta\eta}-M'_{\eta}G'_{\eta}
+\frac{1}{2 T'^{2}}E'_{\eta}\overline{E'}_{\eta}
=\frac{1}{2T'^{2}}\{\mu_{2}\nu_{2}M^{2}_{\eta}+(\mu_{2}\nu_{3}
+\mu_{3}\nu_{2})M_{\eta}E_{\eta}\nonumber\\&&+(\mu_{2}\nu_{4}
+\mu_{4}\nu_{2})M_{\eta}\overline{E}_{\eta}+\mu_{3}\nu_{3}E^{2}_{\eta}
+\mu_{4}\nu_{4}\overline{E}^{2}_{\eta}\}+\{-\frac{\zeta_{1}}{2T^{2}}\nonumber\\
&&+\frac{\mu_{3}\nu_{4}+\mu_{4}\nu_{3}}{2T'^{2}}\}E_{\eta}\overline{E}_{\eta}
+\{\zeta_{0G}+\zeta_{0G'}\zeta_{1}+\zeta_{1\eta}+\zeta_{1G'}\zeta_{0}
+\zeta_{1M'}\lambda_{0}+\zeta_{1E'}\mu_{0}\nonumber\\&&+\zeta_{1\overline{E'}}\nu_{0}
-\lambda_{0}\zeta_{1}\}G_{\eta}+ ... =0 \label{316}
\end{eqnarray}
\begin{eqnarray}
&&M'_{\eta\xi}-G'_{\eta}G'_{\xi}+\frac{1}{4T'^{2}}(E'_{\xi}\overline{E'}_{\eta}
+\overline{E'}_{\xi}E'_{\eta})=(\lambda_{2}-\zeta^{2}_{1})G_{\xi}G_{\eta}
-\zeta_{1}\overline{\zeta_{0}}G_{\eta}\nonumber\\
&&+\frac{1}{2T'^{2}}\{(\mu_{3}\overline{\mu_{0}}
+\nu_{3}\overline{\nu_{0}})E_{\eta}+(\mu_{4}\overline{\mu_{0}}
+\nu_{4}\overline{\nu_{0}})\overline{E}_{\eta}\}
-\frac{\lambda_{3}}{2}G_{\eta}E_{\xi}\nonumber\\
&&-\frac{\lambda_{4}}{2}G_{\eta}\overline{E}_{\xi}+ ... =0
\label{317}
\end{eqnarray}
\begin{eqnarray}
&&E'_{\eta\xi}+\frac{1}{2}(G'_{\eta}E'_{\xi}
+G'_{\xi}E'_{\eta})-\frac{1}{T'}E'_{\xi}E'_{\eta}=\mu_{2}G_{\eta}G_{\xi}
+\frac{1}{2}\zeta_{1}\overline{\nu_{0}}G_{\eta}\nonumber \\
&&+(\mu_{3\xi}+\frac{\overline{\zeta_{0}}\mu_{3}}{2}
-\frac{\mu_{3}\overline{\nu_{0}}}{T'})E_{\eta}+(\mu_{4\xi}
+\frac{\overline{\zeta_{0}}\mu_{4}}{2}
-\frac{\mu_{4}\overline{\nu_{0}}}{T'}){\overline E}_{\eta}+(-\frac{\mu_{3}}{2}\nonumber \\
&&+\frac{\zeta_{1}\overline{\nu_{4}}}{2})G_{\eta}E_{\xi}+(-\frac{\mu_{4}}{2}
+\frac{\zeta_{1}\overline{\nu_{3}}}{2})G_{\eta}\overline{E}_{\xi}
+\frac{1}{2}\zeta_{1}\overline{\nu_{2}}G_{\eta}M_{\xi}+ ... =0
\label{318}
\end{eqnarray}
\begin{eqnarray}
&&E'_{\xi\eta}+\frac{1}{2}(G'_{\eta}E'_{\xi}
+G'_{\xi}E'_{\eta})-\frac{1}{T'}E'_{\xi}E'_{\eta}
=\overline{\nu_{2}}G_{\eta}G_{\xi}
+\frac{1}{2}\zeta_{1}\mu_{0}G_{\xi}\nonumber \\
&&+(\overline{\nu_{4}}_{\eta}+\frac{\zeta_{0}\overline{\nu_{4}}}{2}
-\frac{\mu_{0}\overline{\nu_{4}}}{T'})E_{\xi}+(\overline{\nu}_{3\eta}
+\frac{\zeta_{0}\overline{\nu_{3}}}{2}
-\frac{\mu_{0}\overline{\nu_{3}}}{T'})\overline{E}_{\xi}
+(-\frac{\overline{\nu_{4}}}{2}\nonumber \\
&&+\frac{\zeta_{1}\mu_{3}}{2})G_{\xi}E_{\eta}
+(-\frac{\overline{\nu_{3}}}{2}
+\frac{\zeta_{1}\mu_{4}}{2})G_{\xi}{
\overline{E}}_{\eta}
+\frac{1}{2}\zeta_{1}\mu_{2}G_{\xi}M_{\eta}+ ... =0
\label{319}
\end{eqnarray}

We equate to zero the coefficients of the linearly independent terms
$ M^{2}_{\eta} $, $ M_{\eta}E_{\eta} $, $ M_{\eta}\overline{E_{\eta}} $, ...
and solve the resulting equations. Then we get :

Case 1
\begin{equation}
\zeta_{1}=\lambda_{2}=\lambda_{3}=\lambda_{4}=\mu_{2}=\mu_{3}=\mu_{4}
=\nu_{2}=\nu_{3}=\nu_{4}=0 \label{320}
\end{equation}

Case 2
\begin{eqnarray}
&&\zeta_{0}=\lambda_{0}=\lambda_{3}=\lambda_{4}=\mu_{0}=\mu_{2}=\mu_{4}
=\nu_{0}=\nu_{2}=\nu_{3}=0 \label{321}\\
&&\zeta_{1}=\pm1, \; \lambda_{2}=1, \;
\nu_{4}=\pm\overline{\mu_{3}}, \; \mu_{3\xi}=\mu_{3\eta}=0
\label{322}\\
&&\mu_{3}\overline{\mu_{3}}=\frac{T'^{2}}{T^{2}} \label{323}
\end{eqnarray}

Case 3
\begin{eqnarray}
&&\zeta_{0}=\lambda_{0}=\lambda_{3}=\lambda_{4}=\mu_{0}=\mu_{2}=\mu_{3}
=\nu_{0}=\nu_{2}=\nu_{4}=0 \label{324}\\
&&\zeta_{1}=\pm1, \; \lambda_{2}=1, \;
\nu_{3}=\pm\overline{\mu_{4}}, \; \mu_{4\xi}=\mu_{4\eta}=0 \label{325}\\
&&\mu_{4}\overline{\mu_{4}}=\frac{T'^{2}}{T^{2}} \label{326}
\end{eqnarray}

We can further show that the expressions with the minus sign in
Eqs (\ref{322}) and (\ref{325}) should be dropped. Indeed we get
for this sign and Case 2
\begin{equation}
E'_{\eta\xi}-E'_{\xi\eta}=(-\mu_{3}+\mu_{3G})G_{\eta}E_{\xi}+...=0
\label{327}
\end{equation}
The above implies the relation $ \mu_{3G}-\mu_{3}=0 $, which is
not consistent with Eq (\ref{323}). Similarly we proceed in Case 3
.
Therefore we find the following BTs :

 Case 1
\begin{equation}
G'_{\eta}=\zeta_{0}, \; M'_{\eta}=\lambda_{0}, \;
E'_{\eta}=\mu_{0}, \; \overline{E'}_{\eta}=\nu_{0} \label{328}
\end{equation}
This case is not interesting.

Case 2
\begin{eqnarray}
&&G'_{\eta}=G_{\eta}, \; M'_{\eta}=M_{\eta},\;
E'_{\eta}=\mu_{3}E_{\eta}, \;
\overline{E'}_{\eta}=\overline{\mu_{3}}\overline{E}_{\eta}
\label{329}\\
&&\mu_{3\xi}=\mu_{3\eta}=0, \;
\mu_{3}\overline{\mu_{3}}=\frac{T'^{2}}{T^{2}} \label{330}
\end{eqnarray}

Case 3
\begin{eqnarray}
&&G'_{\eta}=G_{\eta}, \; M'_{\eta}=M_{\eta}, \;
E'_{\eta}=\mu_{4}\overline{E}_{\eta}, \; \overline{E'}_{\eta}
=\overline{\mu_{4}}E_{\eta} \label{331} \\
&&\mu_{4\xi}=\mu_{4\eta}=0, \;
\mu_{4}\overline{\mu_{4}}=\frac{T'^{2}}{T^{2}} \label{332}
\end{eqnarray}
We shall consider Cases 2 and 3 only.

We can easily check that the BTs of Eqs (\ref{329}) - (\ref{330})
and also the BTs of Eqs (\ref{331}) - (\ref{332}) satisfy Eqs
(\ref{31}) and (\ref{32}). Therefore we shall determine $ \mu_{3}
$ and $ \mu_{4} $ by requiring Eq (\ref{33}) to be satisfied.
Substituting expressions (\ref{329})  into Eq (\ref{33}), using Eq
(\ref{33}) to eliminate $ E_{\xi\eta} $ from the resulting
equation and equating to zero the coefficients of the linearly
independent terms we get
\begin{eqnarray}
&&\mu_{3G}=\mu_{3M}=0 \label{333} \\
&&\mu_{3E}+\mu_{3}\mu_{3E'}+\frac{1}{T}\mu_{3}-\frac{1}{T'}\mu^{2}_{3}=0
\label{334} \\
&&\mu_{3\overline{E}}+\overline{\mu_{3}}\mu_{3\overline{E'}}=0
\label{335}
\end{eqnarray}
If we write
\begin{equation}
\mu_{3}=\frac{T'}{T}v, \; \mu_{4}=\frac{T'}{T}w \label{336}
\end{equation}
we get from Eqs (\ref{330}) and (\ref{332})
\begin{equation}
v\overline{v}=1, \; w\overline{w}=1  \label{337}
\end{equation}
Substituting expression (\ref{336}) for $ \mu_{3} $ into Eqs (\ref{330})
and (\ref{333}) - (\ref{335}) we get
\begin{eqnarray}
&&v_{\xi}=v_{\eta}=v_{G}=v_{M}=0 \label{338}\\
&&Tv_{E}+T'vv_{E'}+\frac{1}{2}(v-v^{2})=0 \label{339}\\
&&Tv_{\overline E} -T'v{\overline v}_{\overline E'}+\frac{1}{2}(1-v)=0
\label{340}
\end{eqnarray}

Any complex number $ v $ satisfying the first of (\ref{337}) can be written
in the form
\begin{equation}
v=\frac{x+i}{x-i}, \; x \mbox{ : real} \label{341}
\end{equation}
Substituting expression (\ref{341}) into Eqs (\ref{338}) -
(\ref{340}) we get in real variables
\begin{eqnarray}
&&x_{\xi}=x_{\eta}=x_{G}=x_{M}=0  \label{342}\\
&&Txx_{T}-Tx_{\phi}+T'xx_{T'}+T'x_{\phi'}+x^{2}+1=0 \label{343}\\
&&Tx_{T}+Txx_{\phi}-T'x_{T'}+T'xx_{\phi'}=0 \label{344}\\
&&T(x^{2}+1)x_{T}+T'(x^{2}-1)x_{T'}+2T'xx_{\phi'}+x(x^{2}+1)=0
\label{345}\\
&&T(x^{2}+1)x_{\phi}-2T'xx_{T'}+T'(x^{2}-1)x_{\phi'}-x^{2}-1=0
\label{346}
\end{eqnarray}But multiplying (\ref{343}) by $ x $ and adding to
it Eq (\ref{344}) we get Eq (\ref{345}), while multiplying
(\ref{344}) by $ x $ and subtracting from it Eq (\ref{343}) we get
Eq (\ref{346}). Therefore if Eqs (\ref{343}) and (\ref{344}) are
satisfied Eqs (\ref{345}) and (\ref{346} are also satisfied and
can be omitted. Also the integrability condition $
E'_{\xi\eta}-E'_{\eta\xi}=0 $ leads to Eq (\ref{335}). Therefore
the only relations $ x $ should satisfy are Eqs (\ref{342}) -
(\ref{344}).

If we start from the BTs of Eqs (\ref{331}) - (\ref{332}), write
\begin{equation}
w=\frac{y+i}{y-i}, \; y \mbox{ : real} \label{347}
\end{equation}
and proceed as before we find that $ y $ must satisfy the
relations
\begin{eqnarray}
&&y_{\xi}=y_{\eta}=y_{G}=y_{M}=0 \label{348}\\
&&Tyy_{T}+Ty_{\phi}+T'yy_{T'}+T'y_{\phi'}+y^{2}+1=0 \label{349}\\
&&Ty_{T}-Tyy_{\phi}-T'y_{T'}+T'yy_{\phi'}=0 \label{350}
\end{eqnarray}
which are obtained from Eqs (\ref{342}) - (\ref{344}) if we make
the replacement
\begin{equation}
x\rightarrow y, \; \phi\rightarrow -\phi \label{351}
\end{equation}

We shall find the common solution of Eqs (\ref{343}) and
(\ref{344}). To do that we find separately the general solution of
(\ref{343}) and of (\ref{344}) and then the most general common
solution. To solve (\ref{343}) we find first the solution of the
system
\begin{equation}
\frac{dT}{Tx}=-\frac{d\phi}{T}=\frac{dT'}{T'x}=\frac{d\phi'}{T'}
=-\frac{dx}{x^{2}+1} \label{352}
\end{equation}
Solving this system we find that the general solution of
(\ref{343}) is given by
\begin{equation}
H_{1}\{\frac{T'}{T}, \; T\sqrt{x^{2}+1}, \; Tx-\phi, \;
T'x+\phi'\}=0 \label{353}
\end{equation}
where $ H_{1} $ is an arbitrary function of its arguments. Similarly we
find that the general solution of (\ref{344}) is given by
\begin{equation}
H_{2}\{TT', \; x, \; Tx-\phi, \; T'x+\phi'\}=0 \label{354}
\end{equation}
where $ H_{2} $ is again an arbitrary function of its arguments. The
common solution of (\ref{343}) and (\ref{344}) can be expressed by
an arbitrary function $ H $ whose arguments can be constructed
from the arguments of $ H_{1} $ and also from the arguments of
$ H_{2} $ and is given by
\begin{equation}
H\{TT'(x^{2}+1), \; Tx-\phi, \; T'x+\phi'\}=0 \label{355}
\end{equation}

The above expression gives the general solution of the system in
the sense that no arbitrary function of four arguments can express
its solution. Indeed the general solution of Eqs (\ref{343}) and
(\ref{344}) are given by the arbitrary functions
\begin{eqnarray}
&&H'_{1}\{\frac{T'}{T}, \; TT'(x^{2}+1), \; Tx-\phi, \;
T'x+\phi'\}=0 \label{356}\\ &&H'_{2}\{TT', \; TT'(x^{2}+1), \;
Tx-\phi, \; T'x+\phi'\}=0 \label{357}
\end{eqnarray}
The first arguments $ \frac{T'}{T} $ and $ TT' $ of $ H'_{1} $ and
$ H'_{2} $ cannot be replaced by a common expression which is
constructed from the four arguments of $ H'_{1} $ and
independently from the four arguments of $ H'_{2} $ . We can also
say that if this was possible then (\ref{343}) and (\ref{344}),
which are different equations, would have the same general
solution. But this is impossible and the proof is completed.

Using Eq (\ref{341}) we can eliminate $ x $ from the arguments of $ H $.
We find
\begin{eqnarray}
&&A\equiv
-i(xT-\phi)=E+\frac{E+\overline{E}}{v-1} \label{358}\\
&&B\equiv
-i(xT'+\phi')=\overline{E'}+\frac{E'+\overline{E'}}{v-1}
\label{359}\\
&&D\equiv
TT'(x^{2}+1)-(xT-\phi)(xT'+\phi')=E\overline{E'}+\frac{E\overline{E'}
-\overline{E}E'}{v-1} \label{360}
\end{eqnarray}
and expression (\ref{355}) becomes
\begin{equation}
H(A, \: B, \: D)=H\{E+\frac{E+\overline{E}}{v-1}, \;
\overline{E'}+\frac{E'+\overline{E'}}{v-1}, \; E\overline{E'}+
\frac{E\overline{E'}-\overline{E}E'}{v-1}\}=0 \label{361}
\end{equation}
Also the BTs of Eqs (\ref{329}) - (\ref{330}) for
the variables $ E $ and $ E' $ become
\begin{equation}
E'_{\alpha}=\frac{E'+\overline{E'}}{E+\overline{E}}\,v\,E_{\alpha}, \;
\overline{E'}_{\alpha}=\frac{E'+\overline{E'}}{E
+E'}\,\frac{1}{v}\,\overline{E}_{\alpha},\; \alpha=\xi, \; \eta \label{362}
\end{equation}
Expressions (\ref{362}) with $ v $ given by Eq (\ref{361}) are the BTs
of Case 2.

To solve Eqs (\ref{362}) we proceed as follows \cite{K10} : We get
if we use these equations
\begin{eqnarray}
&&\partial_{\alpha}A=-\frac{E+E'}{(v-1)^{2}}\{v_{\alpha}-\frac{v-1}{E
+\overline E}(vE_{\alpha}+{\overline E}_{\alpha})\} \label{363}\\
&&\partial_{\alpha}B=-\frac{E'+\overline{E'}}{(v-1)^{2}}
\{v_{\alpha}-\frac{v-1}{E +\overline{E}}(vE_{\alpha}+{\overline
E}_{\alpha})\} \label{364}\\
&&\partial_{\alpha}D=-\frac{E\overline{E'}-\overline{E}E'}{(v-1)^{2}}
\{v_{\alpha}-\frac{v-1}{E +\overline{E}}(vE_{\alpha}+{\overline
E}_{\alpha})\} \label{365}
\end{eqnarray}
and therefore we have
\begin{eqnarray}
\partial_{\alpha}H(A, \: B, \: D)=-\frac{1}{(v-1)^{2}}\{(E
+\overline E)\partial_{A}H
+(E'+\overline{E'})\partial_{B}H+(E\overline{E'}\nonumber\\
-{\overline E}E')\partial_{D}H\}
\{v_{\alpha}-\frac{v-1}{E
+\overline{E}}(vE_{\alpha}+{\overline E}_{\alpha})\} \label{366}
\end{eqnarray}
Thus at least one of the following relations should hold
\begin{eqnarray}
&&(A-E)\partial_{A}H+(B-\overline{E'})\partial_{B}H+(D
-E\overline{E'})\partial_{D}H=0 \label{367}\\
&&v_{\alpha}-\frac{v-1}{E
+\overline{E}}(vE_{\alpha}+{\overline E}_{\alpha})=0  \label{368}
\end{eqnarray}
where in writing (\ref{367}) we have used Eqs (\ref{358}) -
(\ref{360}).

Assume that Eq (\ref{367}) holds, which means that $ v $ coming from
it satisfies Eqs (\ref{343}) and (\ref{344}). Since $ E $ ,
$\overline{E'}$ and $E\overline{E'}$ cannot be obtained from $A$,
 $B$ and $D$ this can happen only if $H$ is of the form $H=H(A)$
 or $H=H(B)$ or $H=H(D)$. But then solving these equations for $A$,
$ B$ or $D$ we get $A=-ic_{1}$, $B=-ic_{2}$ or $D=c_{3}$ respectively
($c_{1}$, $c_{2}$ and $c_{3}$ are constants), which give the relations
\begin{equation}
v=\frac{c_{1}+i\overline E}{c_{1}-iE}, \; v=\frac{c_{2}
+iE'}{c_{2}-i\overline
E'}, \; v=\frac{c_{3}-{\overline E}E'}{c_{3}
-E{\overline E'}} \label{369}
\end{equation}
Since $ v{\overline v} =1 $ the constants $ c_{1} $, $ c_{2} $ and
$ c_{3} $ must be real. However if $ v $ is given by anyone of
(\ref{369}) and $ E' $ and $ \overline E' $ satisfy Eqs
(\ref{362}) the quantity $ v $ satisfies Eq (\ref{368}). Therefore
Eq (\ref{368}) is always satisfied which means that we have found
the BTs (\ref{362}) where the pseudopotential \cite{W11} $ v $ is a
solution of Eq (\ref{368}). In this formalism the pseudopotential
was obtained from the arbitrary function $ H $.

The general solution of Eq (\ref{368}) is
\begin{equation}
v=\frac{c+i\overline E }{c-iE } \label{370}
\end{equation}
where c is an arbitrary real constant. Substituting (\ref{370})
into (\ref{362}) and solving the resulting relation we get
\begin{equation}
E'=\frac{E+ic_{1}}{ic_{2}E+c_{3}} \label{371}
\end{equation}
where $ c_{1} $, $ c_{2} $ and $ c_{3} $ are arbitrary real
constants. The above relation is the transformation of Ehlers
\cite{E5}

To obtain the BTs of Case 3 we have to solve Eqs (\ref{348}) -
(\ref{350}). The general solution of these equations, which comes from
 the expression (\ref{355}) if we make the substitution of Eqs
 (\ref{351}), is given by the relation
 \begin{equation}
 Q\{TT'(y^{2}+1), \; Ty+\phi, \; T'y+\phi'\}=0 \label{372}
 \end{equation}
 where $ Q $ is an arbitrary function of its arguments. Proceeding as
 in the previous case we find that expression (\ref{372}) becomes
 \begin{equation}
 Q\{{\overline E}+\frac{E+\overline E}{w-1}, \; \overline E'+\frac{E'+
 \overline E'}{w-1}, \; {\overline E}\,{\overline E'}
 +\frac{{\overline E}\,{\overline E'}-EE'}{w-1}\}=0
 \label{373}
 \end{equation}
 Also the BTs of Eqs (\ref{331}) and (\ref{332}) can be written in
 the form
 \begin{equation}
 \partial_{\alpha}E'=\frac{E'+\overline E'}{E
 +\overline E}\,w\,\partial_{\alpha}\overline E \mbox{,    }
\partial_{\alpha}\overline E'=\frac{E'+\overline E'}{E
+\overline E}\,\frac{1}{w}\partial_{\alpha}E, \; \alpha=\xi, \; \eta
\label{374}
\end{equation}
The above expressions with $ w $ given by Eq (\ref{373}) are the
BTs of Case 3.

 Proceeding as before we find that
\begin{eqnarray}
w=\frac{c+iE}{c-i\overline E} \label{375}\\
E'=\frac{\overline E+ic_{1}}{ic_{2}\overline E+c_{3}} \label{376}
\end{eqnarray}
where $ c $, $ c_{1} $, $ c_{2} $, $ c_{3} $ are arbitrary real
constants. The BT (\ref{376}) is essentially a trivial
modification of the BT (\ref{371}) since if $ E=T+i\phi $ is a
solution of the system of Eqs (\ref{31}) - (\ref{33}) then $
E=T-i\phi $ is again a solution of the same system.

To find $ p' $ and $ \mu' $ of the new solution we have to use the
relation $ \mu+3p=0 $ and Eqs (\ref{34}), (\ref{329}),
(\ref{331}), (\ref{371}) and (\ref{376}). We find for Case 2 :
\begin{equation}
p'=\frac{c_{1}c_{2}+c_{3}}{c^{2}_{2}}\frac{p}{T^{2}+(\phi
-\frac{c_{3}}{c_{2}})^{2}} \mbox {,    }
\mu'=\frac{c_{1}c_{2}+c_{3}}{c^{2}_{2}}\frac{\mu}{T^{2}+(\phi
-\frac{c_{3}}{c_{2}})^{2}} \label{377}
\end{equation}
while $ p' $ and $ \mu' $ for Case 3 come from the above
expressions if we make the substitution $ \phi \rightarrow -\phi$.

The fact that $ G' $, $ M' $ and $ E' $ of Eqs (\ref{329}) and
(\ref{371}) give a new solution of Eqs (\ref{31}) - (\ref{33}) can
be seen as follows : Under these transformations Eqs (\ref{31})
and (\ref{32}) are invariant, while Eq (\ref{33}) becomes
\begin{equation}
\Lambda(E', \; G')=-\frac{c_{1}c_{2}+c_{3}}{c^{2}_{2}}(E-
i\frac{c_{3}}{c_{2}})^{-3}(E
+i\frac{c_{3}}{c_{2}})^{-1}\Lambda(E, \; G)=0 \label{378}
\end{equation}
i.e. it is also satisfied. Similarly we find that $ G' $, $ M' $
and $ E' $ of Eqs (\ref{331}) and (\ref{376}) give a solution of
Eqs (\ref{31}) - (\ref{33}).


\begin{thebibliography}{99}
\bibitem{M1} R M Miura (Ed), {\em B\"{a}cklund Transformations the
Inverse Scattering Method, Solitons and their Applications }, NSF
Research Workshop on Contact Transformations, Lecture Notes in
Mathematics {\bf{515}} (Springer Verlang, Berlin, 1976)
\bibitem{C2} J Clairin, Ann. Sci. Ecole Norm. Sup.$ 3^{e} $ Ser.
Suppl. {\bf{19}}, 1 (1902)
\bibitem{C3} J Clairin, Ann. Fac. Sci. Univ. Toulouse $ 2^{e} $ Ser. {\bf{5}},
437 (1903)
\bibitem{O4} M Omote and M Wadati, J. Math. Phys. {\bf{22}}, 961 (1981)
\bibitem{E5} J Ehlers, Dissertation, Hamburg (1957)
\bibitem{N6} G Neugebauer, J, Phys, A : Math. Gen. {\bf{12}}, L67 (1979)
\bibitem{H7} B K Harrison, Phys. Rev. Lett. {\bf{41}}, 1197 (1978)
\bibitem{O8} M Omote Y Michihiro and M Wadati, Phys.  Lett. A
{\bf{79}}, 141 (1980)
\bibitem{K9} E Kyriakopoulos {\em Einstein's Field Equations for
the Interior of a Uniformly Rotating, Stationary, Axisymmetric
Perfect Fluid }, NTUA preprint, submitted for publication
\bibitem{K10} E Kyriakopoulos J. Phys. A : Math. Gen. {\bf{20}},
1669 (1987)
\bibitem{W11} H D Wahlquist and F B Estabrook, J. Math. Phys.
{\bf{16}}, 1 {1975}
\end{thebibliography}
\end{document}